\documentclass[apj]{emulateapj}
\bibpunct[; ]{(}{)}{;}{a}{}{;}
\usepackage{apjfonts}

\usepackage{graphicx}
\usepackage{natbib}

\newcommand{\Msolar}{M${_\odot}$\,}

\shorttitle{Why haven't loose globular clusters collapsed yet?}
\shortauthors{De Marchi, Paresce, \& Pulone}

\submitted{ApJ Letters; Received 9 November 2006; Accepted 18 January 2007}

\begin{document}

\title{Why haven't loose globular clusters collapsed yet?}


\author{
Guido De Marchi,\altaffilmark{1}
Francesco Paresce,\altaffilmark{2} and
Luigi Pulone\altaffilmark{3}
}

\altaffiltext{1}{European Space Agency, Space Science Department, Keplerlaan
1, 2200 AG Noordwijk, Netherlands; gdemarchi@rssd.esa.int}

\altaffiltext{2}{INAF, Viale del Parco Mellini 84, 00136 Rome, Italy;
fparesce@inaf.it}
 
\altaffiltext{3}{INAF, Osservatorio Astronomico di Roma, Via di Frascati 33, 
00040 Monte Porzio Catone, Italy; pulone@mporzio.astro.it}

\begin{abstract}  

We report on the discovery of a surprising observed correlation between
the slope of the low-mass stellar global mass function (GMF) of
globular clusters (GCs) and their central concentration parameter
$c=\log(r_t/r_c)$, i.e. the logarithmic ratio of tidal and core radii. 
This result is based on the analysis of a sample of twenty Galactic GCs
with solid GMF measurements from deep HST or VLT data. All the
high-concentration clusters in the sample have a steep GMF, most likely
reflecting their initial mass function (IMF). Conversely,
low-concentration clusters tend to have a flatter GMF implying that
they have lost many stars via evaporation or tidal stripping. No GCs
are found with a flat GMF and high central concentration. This finding
appears counter-intuitive, since the same two-body relaxation mechanism
that causes stars to evaporate and the cluster to eventually dissolve
should also lead to higher central density and possibly core-collapse.
Therefore, more concentrated clusters should have lost proportionately
more stars and have a shallower GMF than low concentration clusters,
contrary to what is observed. It is possible that severely depleted GCs
have also undergone core collapse and have already recovered a normal
radial density profile. It is, however, more likely that GCs with a
flat GMF have a much denser and smaller core than suggested by their
surface brightness profile and may well be undergoing collapse at
present. In either case, we may have so far seriously underestimated
the number of post core-collapse clusters and many may be lurking in
the Milky Way.

\end{abstract}


\section{Introduction}

\defcitealias{pdm00}{PDM}
\defcitealias{dem06}{DMPP}

In recent years an increasing number of globular clusters (GCs) have
been found to be seriously depleted in low-mass  stars ($\lesssim
0.5$\,\Msolar) when compared with high-concentration clusters
(\citealt{pdm00} hereafter \citetalias{pdm00}; \citealt{dem05}). The 
first heavily depleted cluster to be discovered was NGC\,6712 
\citep{dem99, and01}, followed by Pal\,5 \citep{koc04}, NGC\,6218 
(\citealt{dem06}, hereafter \citetalias{dem06}), NGC\,2298 \citep{dem07} 
and NGC\,6838 \citep{pul07}. In all cases, the analysis of the radial 
variation of the directly observed mass function (MF) confirms that 
the relative depletion of low-mass stars is not due to the local 
effect of mass segregation but is a structural property of the global 
mass function (GMF), i.e. the MF of the entire stellar population of 
the GC obtained by model fitting. Hereafter, the GMF is defined as a 
power-law function, where the number of stars $N$ per unit mass $m$ 
follows a relationship of the type $dN/dm \propto m^\alpha$ for $0.3 
< m < 0.8$\,\Msolar.

For some of the clusters (NGC\,6712, Pal\,5, NGC\,6838), this finding
is at least qualitatively consistent with the predicted effect of tidal
stripping caused by the Galaxy as these objects have typically shorter
disruption times ($T_{\rm dis}$) than the average GC according to the
models of \citet{gne97}, \citet{din99} and \citet{bau03}.
Quantitatively, however, the  agreement is poor for all of them, since
the expected tight correlation between $T_{\rm dis}$ and GMF slope
\citep{bau03}, whereby clusters with a higher probability of disruption
should always have a shallower GMF slope, is not observed 
(\citetalias{dem06}; \citealt{dem07}). Rather than an error in the 
models, this is possibly the result of the large uncertainties affecting 
their input parameters, especially the clusters' orbits and exact shape 
of the Galactic potential. (For example, the redetermination of the orbit 
of NGC\,6218 with respect to the Hipparcos reference system has led to a 
value of $T_{\rm dis}$ in better agreement with the cluster's flat GMF
\citepalias{dem06} than that based on the previous, less accurate orbit.)
This situation is unlikely to improve much until the advent of
space astrometry missions like Gaia.

The absence of a clear correlation with the effects of tidal stripping,
on the other hand, could also be due, at least in part, to our
imperfect understanding of the relation between the cluster's GMF and
its fundamental structural parameters and their evolution in time. To
explore this possibility, we have used the available data to search for
possible signs of a more complex situation. We present in this Letter
the results of this study which indeed imply that there is more at work
here than was thought up to now. 

Specifically, we find that there is a systematic trend between the GMF
slope $\alpha$ and the central concentration parameter $c$ (defined as
$\log(r_{\rm t}/r_{\rm c})$ where $r_{\rm t}$ and $r_{\rm c}$ are the
tidal radius and core radius, respectively), in the sense that all five
clusters above with a severely depleted GMF have an intermediate or low
value of $c$. On the contrary, the twelve halo clusters in the sample
studied by \citetalias{pdm00} and \citet{dem05} have typically high central
concentration ($<$$c$$>$ $\simeq 1.9$) and correspondingly much steeper
GMF slopes, typically $\alpha=-1.4$ in the range $0.3 - 0.8$\,\Msolar.
In the following sections, we explore in more detail the origin and
nature  of this trend and its possible explanations.

\section{The sample}

The sample used in this investigation includes all GCs for which
reliable luminosity functions (LFs) and MFs exist from deep HST or VLT
photometry. This includes the twelve halo clusters studied by
\citetalias{pdm00}, the two bulge clusters NGC\,6352 and NGC\,6496 studied
by \citet{pul03} and the five GCs, mentioned above, that have recently
been shown to have a depleted GMF at the low-mass end. Finally, we have
added to the sample NGC\,288, whose LF has been studied by \citet{bel02}
with the HST and also reveals a paucity of low-mass stars. The complete 
sample is listed in Table\,\ref{tab1}, where column (1) gives the cluster 
name, column (2) the bibliographic reference, column (3) the MF index 
$\alpha$ in the mass range $0.3 - 0.8$\,\Msolar, column (4) the value of 
the central concentration parameter $c$ and column (5) the total 
integrated magnitude $M_{\rm V}$, both from Harris (1996). 

\begin{deluxetable}{lcccc}
\tablewidth{245pt}
\tablecaption{Mass function index and central concentration\label{tab1}}
\tablehead{
\colhead{Object} & \colhead{Reference} & \colhead{$\alpha$} & 
\colhead{$c$} & \colhead{$M_V$} }
\startdata
NGC\,104		   & a &  $-1.2$ &  $2.03$ &  $-9.42 $ \\ 
NGC\,288		   & b &  $+0.0$ &  $0.96$ &  $-6.74 $ \\
NGC\,2298\tablenotemark{*} & c &  $+0.5$ &  $1.28$ &  $-6.30 $ \\
Pal\,5\tablenotemark{*}	   & d &  $-0.4$ &  $0.70$ &  $-5.17 $ \\
NGC\,5139		   & a &  $-1.2$ &  $1.61$ &  $-10.29$ \\
NGC\,5272		   & a &  $-1.3$ &  $1.84$ &  $-8.93 $ \\
NGC\,6121\tablenotemark{*} & e &  $-1.0$ &  $1.59$ &  $-7.20 $ \\
NGC\,6218\tablenotemark{*} & f &  $+0.1$ &  $1.29$ &  $-7.32 $ \\
NGC\,6254		   & a &  $-1.1$ &  $1.40$ &  $-7.48 $ \\
NGC\,6341		   & a &  $-1.5$ &  $1.81$ &  $-8.20 $ \\
NGC\,6352		   & g &  $-0.6$ &  $1.10$ &  $-6.48 $ \\
NGC\,6397\tablenotemark{*} & h &  $-1.4$ &  $2.50$ &  $-6.63 $ \\
NGC\,6496		   & g &  $-0.7$ &  $0.70$ &  $-7.23 $ \\
NGC\,6656\tablenotemark{*} & i &  $-1.4$ &  $1.31$ &  $-8.50 $ \\
NGC\,6712\tablenotemark{*} & j &  $+0.9$ &  $0.90$ &  $-7.50 $ \\
NGC\,6752		   & a &  $-1.6$ &  $2.50$ &  $-7.73 $ \\
NGC\,6809		   & a &  $-1.3$ &  $0.76$ &  $-7.55 $ \\
NGC\,6838\tablenotemark{*} & k &  $+0.2$ &  $1.15$ &  $-5.60 $ \\
NGC\,7078\tablenotemark{*} & l &  $-1.9$ &  $2.50$ &  $-9.17 $ \\
NGC\,7099		   & a &  $-1.4$ &  $2.50$ &  $-7.43 $ \\
\enddata
\tablecomments{For clusters marked with an asterisk following the
cluster name the value of $\alpha$ is that of the GMF. For all other
objects, $\alpha$ is the index of the MF measured near the half-mass
radius. Bibliographical references in Column (2) are as follows:
(a) \citealt{pdm00}; (b) \citealt{bel02}; (c) \citealt{dem07}; 
(d) \citealt{koc04}; (e) \citealt{pul99}; (f) \citealt{dem06}; 
(g) \citealt{pul03}; (h) \citealt{dem00}; (i) \citealt{alb02};
(j) \citealt{and01}; (k) \citealt{pul07}; (l) \citealt{pas04}.}
\end{deluxetable}

As for the value of $\alpha$, it has been derived as follows. For those
clusters for which a GMF index exists from multi-mass models (indicated
by an asterisk following the cluster name), that value is used. For all
other clusters, $\alpha$ is the index of the power-law MF that, once
folded through the derivative of the mass--luminosity (M--L)
relationship appropriate for the cluster's metallicity, best fits the
LF measured near the half-mass radius ($r_{\rm hm}$), since there the
MF is expected to closely approach the GMF \citep{dem00}. For clusters
in the sample studied by \citetalias{pdm00}, we adopted the same M--L used
in that  work, while for the remaining objects the M--L relationship of
\citet{bar97} for the appropriate metallicity was used. As pointed out
by \citetalias{pdm00}, since the LF of NGC\,6341 and NGC\,7099
were measured farther out in the cluster, namely at about four times
the half-mass radius, in principle a small positive correction to the
measured index $\alpha$ should be applied to account for the steepening
of the MF with increasing radial distance caused by mass segregation.
This correction amounts to less than $0.2$ dex and is included in the
values listed in Table\,\ref{tab1}. The correction being relatively
small, however, none of our conclusions would change if we were to
ignore it.

\begin{figure}
\centering
\plotone{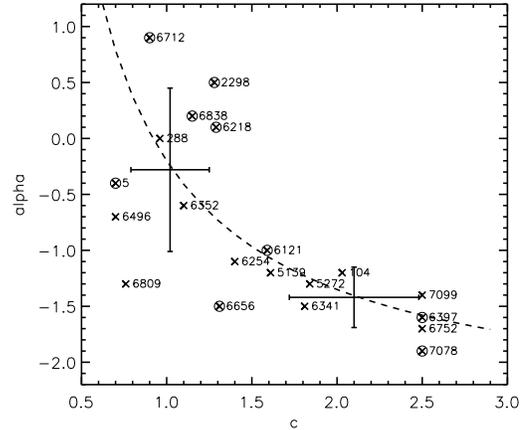}
\caption{Observed trend between MF index $\alpha$ and the central
concentration parameter $c$. Clusters are indicated by their NGC (or
Pal) index number. Objects for which a GMF index is available are
marked with a circled cross. The large crosses indicate the average
and $1\,\sigma$ distribution of $\alpha$ and $c$ for clusters with $c$
values above and below $1.4$. The dashed line is an eye-ball fit to
the distribution.}
\label{fig1}
\end{figure}

In Figure\,\ref{fig1} we show the run of $\alpha$ as a function of $c$
from Table\,\ref{tab1}. It is immediately obvious that there are no 
high-concentration clusters with a shallow GMF. It also appears that a
relatively low concentration is a necessary but not sufficient
condition for a depleted GMF. The median value of $c$ ($1.4$) splits 
the cluster population roughly in two groups, one with lower and one
with higher concentration. The mean GMF index of the first group is
$\alpha=-0.3 \pm 0.7$, while the second has a much tighter
distribution with $\alpha=-1.4 \pm 0.3$. The average values and the
associated $\pm 1\,\sigma$ uncertainties are shown as thick crosses in
Figure\,\ref{fig1}. Due to the large dispersion of $\alpha$ at low
concentration, it is not possible to derive a precise correlation
between $c$ and $\alpha$ over the whole range spanned by these
parameters. On the other hand, the GMF becomes undoubtedly less steep
as $c$ decreases. The relationship $\alpha+2.5=2.3/c$ is a 
simple yet satisfactory eye-ball fit to the distribution (dashed line 
in Figure\,\ref{fig1}).

\section{Discussion}

The result shown in Figure\,1 is very surprising, since there is a
clear absence of centrally concentrated clusters that are depleted in
low-mass stars as one would expect of GCs undergoing tidal disruption,
i.e. objects that should occupy the upper right quadrant of
Figure\,\ref{fig1}. This is contrary to theoretical expectations, since
the same relaxation mechanism that drives a cluster towards higher
central density and, eventually, core collapse is also responsible for
the dissolution of the cluster, via evaporation \citep{spi87}.
Therefore, as a cluster evolves dynamically in the course of its
lifetime, one would expect that its GMF should become shallower at the
low-mass end while the central concentration parameter should increase.
Put differently, severe mass loss should be a sufficient, yet not
necessary, condition for core collapse. In this simple scheme, one
would for instance expect that NGC\,2298 and NGC\,6218, with $\alpha
\simeq 0.3-0.5$ should have a much denser core than NGC\,6397 with
$\alpha \simeq -1.6$. Clearly, this picture is not at all consistent
with the results shown in Figure\,\ref{fig1}.

That cluster concentration somehow creates variations in the IMF is not
a plausible explanation for the observed trend between $c$ and
$\alpha$, since the star formation process could not possibly know
about the future structural properties of the forming cluster. Thus,
unless the high $c$ value of the densest clusters is primordial and a
presently unknown mechanism exists whereby low-mass star formation is
hampered in low density environments, the depletion at the low-mass end
of the GMF must result from mass loss via dynamical evolution. 

It is possible that the clusters that have lost a significant fraction
of their original mass have indeed undergone core collapse and have
already recovered a normal radial density and surface brightness
profile. A large body of theoretical studies exists on the  evolution
of GCs after core collapse \citep{hen75, sug83, goo93, mak96},
according to which they should undergo a homologous re-expansion,
within a thermal timescale, triggered by the energy released by
hardening binaries \citep{hut85}. This process, however, is not
necessarily stable and can lead to re-collapse and subsequent
gravothermal oscillations \citep{sug83, goo87, coh89, mak96} when the
number of stars in the cluster exceeds a threshold that depends on the
mass spectrum. \citet{mur90} show that for a Salpeter IMF the
re-expansion of the core is stable up to $N_{\rm s} \simeq 3 \times
10^5$, a value comparable with the estimated number of objects present
in the most depleted clusters in our sample. It, therefore, appears
plausible that these clusters may have re-expanded after core collapse.
On the other hand, \citet{mur90} predict the observed core radius to
expand over time as $r_{\rm c} \propto t^{0.6}$ and it would take too
long for the core to reach a size comparable to that of the
pre-collapse phase, unless the core shrunk only marginally during
collapse or current models under-estimate the observed size of $r_{\rm
c}$ during the collapse phase of a realistic multi-mass  cluster with a
finite number of stars.

The evaporation timescale increases with increasing total cluster mass
$M_{\rm T}$, so one could possibly understand the trend shown  in
Figure\,\ref{fig1} if clusters of lower central concentration were 
also those with lower mass. Unfortunately, this is not easy to
investigate as less than half the clusters in our sample have a
relatively solid $M_{\rm T}$, based on multi-mass model fitting, while
for the rest $M_{\rm T}$ comes from the total luminosity, i.e. the
integrated magnitude $M_{\rm V}$, by assuming a constant mass-to-light
ratio. The value of $M_{\rm V}$ is shown for each cluster in
Table\,\ref{tab1} and no correlation is found between it and $c$.
Similarly, no correlation exists between $M_{\rm T}$ and $c$ for
clusters with a reliable value of $M_{\rm T}$. For example, NGC\,2298,
NGC\,6218 and NGC\,6656 have masses in the ratio $1:2.4:5.3$, but 
share the same concentration $c \simeq 1.3$. Conversely, NGC\,6397 is 
about half as massive as NGC\,6656, but its $c$ is twice as large.

It may be argued that besides evaporation other tidal mechanisms are 
responsible for the depleted GMFs that we find. In fact, the compressive 
heating that GCs undergo when they cross the Galactic plane (disk 
shocking) or venture close to the Galactic centre (bulge shocking) can 
have a much stronger effect than evaporation, depending on the cluster's 
orbit \citep{agu88, gne97, din99}. However, while bulge and disk shocks 
can cause significant mass loss \citep{spi87,heg03}, they will 
preferentially remove low-mass stars {\em only if} these have
previously been pushed toward the cluster's periphery by mass
segregation \citep{ves97}. Even if for some GCs tidal shocks have been 
as important as (or more important than) internal two-body relaxation 
in determining the mass loss rate, the observed trend is still puzzling 
since in any case tidal shocks should accelerate the evolution of a 
cluster toward higher central density and core collapse
\citep[see][]{spi73, che86, spi87}. Only initially very loose GCs ($c 
\la 0.5$) are expected to quickly dissolve, via mass loss due to
stellar evolution, before reaching core collapse \citep{che90, fuk95}. 

However, Figure\,\ref{fig1} shows that also clusters of intermediate
concentration may undergo severe mass loss without necessarily showing
signs of core collapse for as long as a Hubble time. This behaviour is
in principle consistent with the results of Fokker--Planck and N-body
calculations of realistic clusters, including the effect of stellar
evolution and two-body relaxation. \citet{che90} and \citet{tak00} have
investigated the temporal evolution of $c$ and $\alpha$ for various
initial conditions (total mass, concentration, relaxation time and IMF
index). Their models suggest that, unless the IMF is very steep
($\alpha \simeq -3.5$), energy equipartition and mass segregation will
initially drive the cluster towards lower values of the central
density, mainly because the cluster shrinks (and therefore stars are
lost) more quickly than the core can contract. Eventually the cluster
undergoes core collapse, but how long this takes and whether the
increase in the central concentration $c$ can measurably affect the
surface brightness profile depends on the initial conditions and in
particular on the IMF slope. If the latter is very shallow ($\alpha
\simeq -1.5$), stellar evolution may remove enough mass from the
cluster so that its core is reduced to just a few stars. Even in the
deepest collapse phase, no central cusp would be visible in the 
surface brightness profile. 

The predictions of \citet{che90} and \citet{tak00} are difficult to
compare directly to the actual data, since the measured value of $c$,
based on the surface brightness profile dominated by red-giant stars,
is not a good tracer of the true cluster density. Nevertheless, the
overall picture seems compatible with Figure\,\ref{fig1} and their
findings allow us to put forth the following hypotheses to explain the
presence of clusters with a shallow GMF and low $c$ and the absence of
objects with a shallow GMF and high $c$.

The dashed line in Figure\,1 approximately traces the evolutionary path
of GCs from their birth towards two opposite directions of increasing
or decreasing concentration. Clusters born with sufficiently high
concentration ($c \ga 1.5$) evolve towards core collapse. Mass loss can
be important via stellar evolution in the first $\sim 1$\,Gyr, and to a
lesser extent via evaporation or tidal stripping throughout the life of
the cluster, but the GMF at any time does not depart significantly from
the IMF. Clusters with $c \la 1.5$ at birth also evolve towards core 
collapse, but mass loss via stellar evolution and, most importantly, 
via relaxation and tidal stripping proceeds faster, particularly if the
orbit has a short perigalactic distance or frequent disk crossings.
Therefore, as the tidal boundary shrinks and the cluster loses
preferentially low-mass stars, the GMF progressively flattens. This
speeds up energy equipartition, but $c$ still decreases, since the
tidal radius shrinks more quickly than the luminous core radius
(although the central density, particularly that of heavy remnants, is
increasing). These clusters can eventually undergo core collapse, but
this might only affect a few stars in the core, thereby making it
observationally hard to detect.

An alternative possibility is that there were IMF variations. Some
clusters with shallow IMF have undergone severe stellar mass-loss and
have therefore expanded considerably. This has led to a lower $c$ and a
shallower GMF (larger $\alpha$) because these systems were more prone
to tidal stripping. Most of these clusters have already disrupted but
some survive for a long time in a state of low $c$ and large $\alpha$
before collapse occurs. Clusters with a steeper IMF, on the other hand,
have proceeded normally to core collapse. 

The problem with this alternative possibility is the absence of GCs
with a shallow GMF slope and high concentration ($c > 1.5$). Since high
initial concentration should lead to a rapid collapse phase
\citep{che90} and no mechanism is known that could steepen the GMF over
time, originally massive clusters with a shallow IMF and high $c$
should still be visible in the ``zone of avoidance'' (upper right
corner) of Figure\,\ref{fig1}. Admittedly, our sample is relatively
small and we cannot exclude that clusters exist in this region of the
parameter space. In particular, GCs of high concentration ($c \ga 2$)
and low total luminosity ($M_{\mathrm V} \ga -6$) are potential 
candidates and their GMF should be investigated. 

Conversely, clusters with low concentration and steep IMF can have
formed, NGC\,6809 being a good candidate. Objects like these should
follow the general evolution towards lower $\alpha$ and possibly lower
$c$. The exact balance between decrease in $\alpha$ and in $c$ should
depend on the initial conditions, and particularly the initial mass
(which together with the tidal radius, set by the Galactic potential,
defines the relaxation time).

Therefore, until GCs are found in the ``zone of avoidance'' of
Figure\,\ref{fig1}, it seems plausible that opposite evolutionary paths
exist in the $c$-vs-$\alpha$ plane for clusters born with different
central concentration and/or on different orbits, but that the IMF was
the same or very similar for all of them. At low masses, the IMF of GCs
must approach the steepest GMFs in our sample, while at higher masses
it cannot be much shallower than Salpeter without mass loss via stellar
evolution causing the rapid disruption of the cluster \citep{che90,
tak00}. We presently have no direct measurements of the GC IMF above
$0.8$\,\Msolar, but a value of $\alpha\simeq -2$ is the preferred
outcome of multi-mass Michie--King models (\citetalias{pdm00}; 
\citetalias{dem06}; \citealt{dem07}). In this sense, the tapered 
power-law distribution proposed by \citet{dem05} remains a viable 
hypothesis for the IMF of GCs.

\section{Conclusions}

We have discovered an empirical correlation between the central
concentration and the GMF slope of GCs, whereby only loose clusters 
have a shallow GMF. A low value of the central concentration seems,
therefore, a necessary condition for extensive mass loss leading to
cluster disruption. Although it is possible that GCs formed with a
certain degree of mass segregation and that some low-mass stars may
have been lost due to tidal truncation before two-body relaxation could
act upon them, all of the depleted clusters that we have studied have a
dynamical structure consistent with their being in a condition of
energy equipartition (\citealt{and01}; \citetalias{dem06}; 
\citealt{dem07, pul07}). Therefore, the observed trend between $\alpha$ 
and $c$ can only be understood if either the depleted clusters have 
undergone collapse and have subsequently rebounded, or if they are 
proceeding unnoticed towards core collapse or have already reached it 
without showing it. In either case, this means that the central 
concentration parameter $c$ derived from the surface brightness profile 
is not a good tracer of a cluster's true central density or dynamical 
state. Our current estimate of the fraction of post-core collapse 
clusters may therefore need a complete revision as a large number of 
them may be lurking in the Milky Way. A reliable assessment of a 
cluster's dynamical state requires the study of the complete radial 
variation of its stellar MF.

\acknowledgements

We thank an anonymous referee whose thorough comments and suggestions
have helped us to considerably improve the presentation of this paper.
It is a pleasure to thank Torsten B\"oker and Andres Jord\'an for very
useful discussions.

\end{document}